# Mathematics as the Language of Physics.

J. Dunning-Davies,
Department of Physics,
University of Hull,
Hull HU6 7RX,
England.

Email: j.dunning-davies@hull.ac.uk

**Abstract.**

Courses in mathematical methods for physics students are not known for including too much in the way of mathematical rigour and, in some ways, understandably so. However, the conditions under which some quite commonly used mathematical expressions are valid can be of great importance in some physical circumstances. Here one such expression, which figures frequently in the manipulations leading to the isothermal compressibility appearing in formulae, is examined as an illustrative example.

Mathematics is undoubtedly a subject which deserves to be studied in its own right. As such it provides a superb exercise for the mind and can easily be seen to lead to people possessed of a flexible thinking which may be applied in myriad areas. However, mathematics also has a vital role to play as the language of physics. As such, the part it plays must be subservient to the physics but, having said that, its role is no less important than that it enjoys when it is studied as a subject in its own right. In both roles the conditions affecting the validity of results are equally important and it is unfortunate that this aspect of mathematics – so important to the pure mathematician – should appear to be overlooked on so many occasions within physics. One mathematical result which illustrates this to perfection comes from the realms of partial differentiation.

Rather than use general mathematical symbols as variables, consider the case of an ideal gas for which the equation of state allows the pressure, *p*, to be written as a function of the volume, *V,* and absolute temperature, *T*; that is,
$$p = p(V, T).$$
For this function
$$dp = \left(\frac{\partial p}{\partial T}\right)_V dT + \left(\frac{\partial p}{\partial V}\right)_T dV.$$
If *V* is now considered as a function of *p* and *T*, it follows that
$$dp = \left(\frac{\partial p}{\partial T}\right)_V dT + \left(\frac{\partial p}{\partial V}\right)_T \left\{\left(\frac{\partial V}{\partial p}\right)_T dp + \left(\frac{\partial V}{\partial T}\right)_p dT\right\}.$$
Comparing coefficients of *dT* on both sides of this equation leads to
$$0 = \left(\frac{\partial p}{\partial T}\right)_V + \left(\frac{\partial p}{\partial V}\right)_T \left(\frac{\partial V}{\partial T}\right)_p$$
or, more familiarly
$$-1 = \left(\frac{\partial p}{\partial V}\right)_T \left(\frac{\partial V}{\partial T}\right)_p \left(\frac{\partial T}{\partial p}\right)_V. \tag{1}$$
This result receives quite widespread use in various areas of physics, particularly in thermodynamics and statistical mechanics. However, it is vitally important to note that the result is valid subject to some simple straightforward conditions – mathematical conditions which, in this case, translate easily into obvious physical requirements. The derivation is clearly dependent on the pressure, *p*, being a definite function of both the volume, *V*, and the absolute temperature, *T*. This condition, simple though it seems, is not always met in physical situations. For example, as has been mentioned previously[1], the pressure of a gas becomes independent of its volume below the condensation temperature[2] and so, below that temperature, result (1) would not be applicable. If, in fact, the result is used, it can lead to incorrect conclusions as has been shown[1] quite clearly for the case of an ideal relativistic Bose gas.

The case of the Bose gas, to which reference is made, raises questions immediately when use is made of the isothermal compressibility in examinations of a system. Again, as has been shown[1], from the definition of the grand partition function, it follows directly that the mean square relative fluctuation in number of particles is given by

$$\frac{\overline{(N-\overline{N})^2}}{(\overline{N})^2} = \frac{kT}{N^2}\left(\frac{\partial N}{\partial \mu}\right)_{V,T}. \quad (2)$$

By using (1) above together with a consequence of the Gibbs-Duhem equation[3], it follows that

$$\frac{\overline{(N-\overline{N})^2}}{(\overline{N})^2} = -\frac{kT}{V^2}\frac{1}{(\partial p/\partial V)_{N,T}} = \frac{k}{V}\mathrm{K}_T, \quad (3)$$

where $\mathrm{K}_T$ is the isothermal compressibility.

As has been shown previously[1], using (3) to evaluate the mean square relative fluctuations in number of particles below the condensation temperature can lead to errors. On the other hand, (2) is a generally valid result. This example is cited merely to draw attention to a very real problem within physics and that is that, quite often, results are derived or quoted with no mention of their ranges of validity. If such ranges are not mentioned, the implication must be that the results are valid quite generally. This lack of mention of ranges of validity should pose no problem to the professional physicist although, quite understandably, it does on occasions. However, for the student the position is left unclear. Students do not have the experience or expertise to realise that some results are valid quite generally, while others are not. Frequently, though not always, mathematical rigour is not a notion overstressed in undergraduate physics courses. Hence, it becomes more and more important to make it absolutely clear when results are valid generally, and when not.

As indicated already, one particular area of concern surrounds results dependent on the expression for the isothermal compressibility. Firstly, its definition; is it[4]

$$\mathrm{K}_T = -\frac{1}{V}\frac{1}{(\partial p/\partial V)_{N,T}}$$

or[5]

$$\mathrm{K}_T = -\frac{1}{V}\left(\frac{\partial V}{\partial p}\right)_{N,T} ?$$

Firstly, the conditions under which these two expressions are actually the same need to be noted and one is that $(\partial p/\partial V)_{N,T}$ be both finite and non-zero. Also, as is seen, the important point arising when the pressure becomes independent of the volume is more readily visible via the first of these definitions; it is somewhat masked in the second. However, as was seen to occur when considering mean square fluctuations in number of particles, the isothermal compressibility often occurs in forms of formulae modified by use of equations of the form of (1) above. It doesn't regularly appear in the initial forms of formulae as is illustrated by the example mentioned above and by considering the equation giving the difference between the constant pressure and constant volume heat capacities:

If the entropy is assumed to be a function of both the absolute temperature and the volume, it follows that

$$dS = \left(\frac{\partial S}{\partial T}\right)_V dT + \left(\frac{\partial S}{\partial V}\right)_T dV,$$

from which it is readily seen that

$$\left(\frac{\partial S}{\partial T}\right)_p = \left(\frac{\partial S}{\partial T}\right)_V + \left(\frac{\partial S}{\partial V}\right)_T \left(\frac{\partial V}{\partial T}\right)_p.$$

By using a Maxwell relation, this is seen to be equivalent to

$$\left(\frac{\partial S}{\partial T}\right)_p = \left(\frac{\partial S}{\partial T}\right)_V + \left(\frac{\partial p}{\partial T}\right)_V \left(\frac{\partial V}{\partial T}\right)_p.$$

By using (1) above, this becomes

$$\left(\frac{\partial S}{\partial T}\right)_p = \left(\frac{\partial S}{\partial T}\right)_V - \left(\frac{\partial p}{\partial V}\right)_T \left[\left(\frac{\partial V}{\partial T}\right)_p\right]^2,$$

which may be written

$$C_p = C_V + VT\frac{\alpha_p^2}{K_T},$$

where $\alpha_p = \frac{1}{V}\left(\frac{\partial V}{\partial T}\right)_p$ is the coefficient of volume expansion at constant pressure.

Yet again, the isothermal compressibility appears in an expression following use of equation (1) and so, in situations occurring below the condensation temperature where the pressure becomes independent of the volume, great care must be taken when considering use of this final form of the result since, as has been mentioned already but cannot be overstressed, below the condensation temperature the pressure becomes independent of the volume and so use of (1) is *not* allowed and expressions derived using it must be invalid for that temperature region.

Attention has been restricted here solely to problems arising from use of equation (1) and, in particular, when that relation is used to introduce the isothermal compressibility into equations. The reason for doing that is clear in that the isothermal compressibility is a quantity whose value may more readily be determined than the values of those quantities it has replaced. However, the above illustrates the need for extreme caution when using results involving this quantity. Nevertheless, this is merely an example; the need for caution when utilising mathematical results is always present. The one comforting thought for the physicist is that, when some of the mathematical results become invalid, the physical situation existing is often unusual also – in the above case of the Bose gas, a change of phase has occurred and it is obvious from everyday examples, such as the transitions between the various states of water, that peculiar things are happening physically at phase transitions.

**References.**

1.  J. Dunning-Davies; 1968, Il Nuovo Cimento, **57B**, 315-320

2.  L. D. Landau and E. M. Lifshitz; 1959, *Statistical Physics,*
        (Pergamon Press, London)

3.  J. Dunning-Davies: 1968, Il Nuovo Cimento, **53B**, 180-181

4.  P. T. Landsberg; 1961, *Thermodynamics with Quantum Statistical Illustrations*,
        (Interscience Publishers, London)

5.   C. J. Adkins; 1968, *Equilibrium Thermodynamics*,
        (McGraw-Hill, London)